\documentclass[iop]{emulateapj}
\usepackage{color}
\usepackage{graphicx}
\usepackage{amsmath}

\shorttitle{Exploring the ``$L$--$\sigma$" relation of HII galaxies}
\shortauthors{Wu, Cao, et al.}

\begin{document}
\title{ Exploring the ``$L$--$\sigma$" relation of HII galaxies and giant extragalactic HII regions acting as standard candles }
\author{Yan Wu\altaffilmark{1}, Shuo Cao\altaffilmark{1$\ast$}, Jia Zhang\altaffilmark{2}, Tonghua Liu\altaffilmark{1$\dag$}, Yuting Liu\altaffilmark{1}, Shuaibo
Geng\altaffilmark{1}, Yujie Lian\altaffilmark{1}}

\altaffiltext{1}{Department of Astronomy, Beijing Normal University,
100875, Beijing, China; \emph{caoshuo@bnu.edu.cn;
liutongh@mail.bnu.edu.cn}} \altaffiltext{2}{School of physics and
Electrical Engineering, Weinan Normal University, Shanxi 714099,
China}

\begin{abstract}

Cosmological applications of HII galaxies (HIIGx) and giant
extragalactic HII regions (GEHR) to construct the Hubble diagram at
higher redshifts require knowledge of the ``$L$--$\sigma$" relation
of the standard candles used. In this paper, we study the properties
of a large sample of 156 sources (25 high-$z$ HII galaxies, 107
local HII galaxies, and 24 giant extragalactic HII regions) compiled
by Terlevich et al.(2015). Using the the cosmological distances
reconstructed through two new cosmology-independent methods, we
investigate the correlation between the H$\beta$ emission-line
luminosity $L$ and ionized-gas velocity dispersion $\sigma$. The
method is based on non-parametric reconstruction using the
measurements of Hubble parameters from cosmic clocks, as well as the
simulated data of gravitational waves from the third-generation
gravitational wave detector (the Einstein Telescope, ET), which can
be considered as standard sirens. Assuming the emission-line
luminosity versus ionized gas velocity dispersion relation, $\log L
($H$\beta) = \alpha \log \sigma($H$\beta)+\kappa$, we find the full
sample provides a tight constraint on the correlation parameters.
However, similar analysis done on three different sub-samples seems
to support the scheme of treating HII galaxies and giant
extragalactic HII regions with distinct strategies. Using the
corrected ``$L$--$\sigma$" relation for the HII observational sample
beyond the current reach of Type Ia supernovae, we obtain a value of
the matter density parameter, $\Omega_{m}=0.314\pm0.054$ (calibrated
with standard clocks) and $\Omega_{m}=0.311\pm0.049$ (calibrated
with standard sirens), in the spatially flat $\Lambda$CDM cosmology.

\end{abstract}


\maketitle

\section{Introduction}

The Hubble diagram, which is directly related to the luminosity
distances, has provided a useful method to probe cosmological
parameters \citep{Riess98,Perlmutter99}. In order to measure the
luminosity distance, we always turn to luminous sources of known (or
standardizable) intrinsic luminosity in the Universe, such as type
Ia supernova (SN Ia) \citep{Cao11,Cao13,Cao15a,Chen15,Qi18} and less
accurate but more luminous gamma-ray bursts (GRB) \citep{Pan15} in
the role of ``standard candles". Powerful HII galaxies and
extragalactic HII regions constitute a population that can be
observed up to very high redshifts, reaching beyond feasible limits
of supernova studies. Indeed, the power of modern cosmology lies in
building up consistency rather than in single and precise
experiments \citep{Biesiada10,Cao15b,Ma19}, which indicates that
every alternative method of restricting cosmological parameters is
desired. It is known that HII galaxies and HII regions of galaxies
could have very similar physical systems
\citep{Melnick1987,JunJie2016}, an outstanding feature of which lies
in the rapidly forming stars surrounded by ionized hydrogen. More
specifically, HII galaxies and HII regions may exhibit
indistinguishable optical spectra, i.e., strong Balmer emission
lines in H$\alpha$ and H$\beta$ due to the hydrogen ionized by the
young massive star clusters
\citep{SS1972,Bergeron1977,TM1981,Kunth2000}.

A well-defined sample of HII galaxies with accurately measured flux
density and the turbulent velocity of the gas could be useful to
test cosmological parameters such as the present-day matter density,
cosmic equation of state, etc. \citep{Siegel2005,Plionis2011}.
Concerning such cosmological applications, the first method used for
this purpose is of statistical nature. Essentially, the idea is to
discuss an important phenomenon that as the mass of the starburst
component increases, both the number of ionized photons and the
turbulent velocity of the gas will increase. Therefore, one may
naturally expect an quantitative relation between the luminosity
$L($H$\beta)$ in H$\beta$ and the ionized gas velocity dispersion
$\sigma$, which has triggered numerous efforts to use HII galaxies
for this purpose \citep{TM1981,Chavez2014}. The first attempt to
determine a possible correlation between the luminosity
$L($H$\beta)$ and profile width for giant HII regions was presented
in \citet{Melnick1979a}, which was then extended to the
luminosity-velocity dispersion relation satisfied by elliptical
galaxies, bulges of spiral galaxies and globular clusters
\citep{TM1981}. It was found that in subsequent analysis
\citep{Melnick1987,Melnick1988} that such ``$L$--$\sigma$" relation,
with small scattering can be used to determine cosmic distances
independently of redshifts. More promising candidates in this
context are HII galaxies (HIIGx) and giant extragalactic HII regions
(GEHR) that can be observed up to very high redshifts. Following the
suggestion proposed by \citet{Pettini1988}, many authors furthermore
confirmed the validity of the ``$L$--$\sigma$" correlation at higher
redshifts \citep{Melnick2000}, which showed that HIIGx and GEHR can
be used as independent distance indicators at $z\sim 3$.

From the original ``$L$--$\sigma$" calibration of a sample of 5
high-$z$ HII galaxies covering the redshift range of $2.1<z<3.4$
\citep{Melnick1988}, in combination with the measurements of flux
density and turbulent gas velocity, \citet{Siegel2005} determined
the best-fit value for the matter density parameter, $\Omega_{m}=
0.21_{-0.12}^{+0.30}$ in the framework of flat $\Lambda$CDM model. A
similar analysis was made by \citet{Plionis2011} concerning the
so-called XCDM cosmology (with constant dark energy equation of
state), using a revised zero-point of the original ``$L$--$\sigma$"
relation \citep{Jarosik2011}. While comparing the results from the
previous ``$L$--$\sigma$" relation, differences in central values of
the best-fit cosmological parameters were also reported: $\Omega_{m}
= 0.22_{-0.04}^{+0.06}$. The possible cosmological application of
these HIIGx and GEHR as a standard candle has been extensively
discussed in the literature
\citep{Fuentes2000,Bosch2002,Telles2003,Siegel2005,Bordalo2011,Plionis2011,Mania2012,Chavez2012,Chavez2014,JunJie2016},
which found that the HII galaxies provides a competitive source of
luminosity distance to probe the acceleration of the Universe. For
instance, more recently, on a new sample of 156 sources compiled by
\citet{Terlevich2015}, \citet{JunJie2016} have studied the
possibility of utilizing HIIGx to carry out comparative studies
between competing cosmologies, such as $\Lambda$CDM and the $Rh=ct$
Universe \citep{Melia07,Melia13}. However, it should be noted that
cosmological application of the HIIGx and GEHR data requires good
knowledge of the ``$L$--$\sigma$" relation of the ``standard
candles" used. One of the major uncertainties was the typical value
of the model parameters ($\alpha$, $\kappa$) of the emission-line
luminosity versus ionized gas velocity dispersion relation, $\log L
($H$\beta) = \alpha \log \sigma($H$\beta)+\kappa$. In order to
obtain cosmological constraints, some authors chose to take $\alpha$
and $\kappa$ as statistical nuisance parameters \citep{JunJie2016}.
One should remember that the nuisance parameters characterizing the
``$L$--$\sigma$" relation introduce considerable uncertainty to the
final determination of other cosmological parameters. Having this in
mind, properties of the HIIGx and GEHR data should be readdressed
with the biggest sample to date (156 combined sources, including 25
high-$z$ HIIGx, 107 local HIIGx, and 24 GEHR) and taking into
account reliable cosmological distance information based on current
precise observations.

This encourages us to improve and develop it further, based on the
newly-compiled sample of Hubble parameter $H(z)$ measurements (which
represents a type of new cosmological standard clock) and the
simulated data of gravitational waves from the third-generation
gravitational wave detector (which can be considered as standard
siren). Compared with the previous works
\citep{Siegel2005,Plionis2011,JunJie2016}, the advantage of this
work is that, we achieve a reasonable and compelling constraints on
the ``$L$--$\sigma$" relation in both the electromagnetic (EM) and
gravitational wave (GW) window, using luminosity distances covering
the HII redshift range derived in two cosmological-model-independent
methods. This paper is organized as follows. We briefly introduce
our methodology and the corresponding observational data (HII,
$H(z)$ and GW) in Section II. Cosmological-model-independent
constraints on the full sample and several sub-samples are presented
in Section III. The cosmological application of the calibrated
``$L$--$\sigma$" relation of HII regions are Section IV. Finally,
the conclusions and discussions are presented in Section IV.

\begin{figure*}[tbp]
\centerline{\includegraphics[scale=0.51]{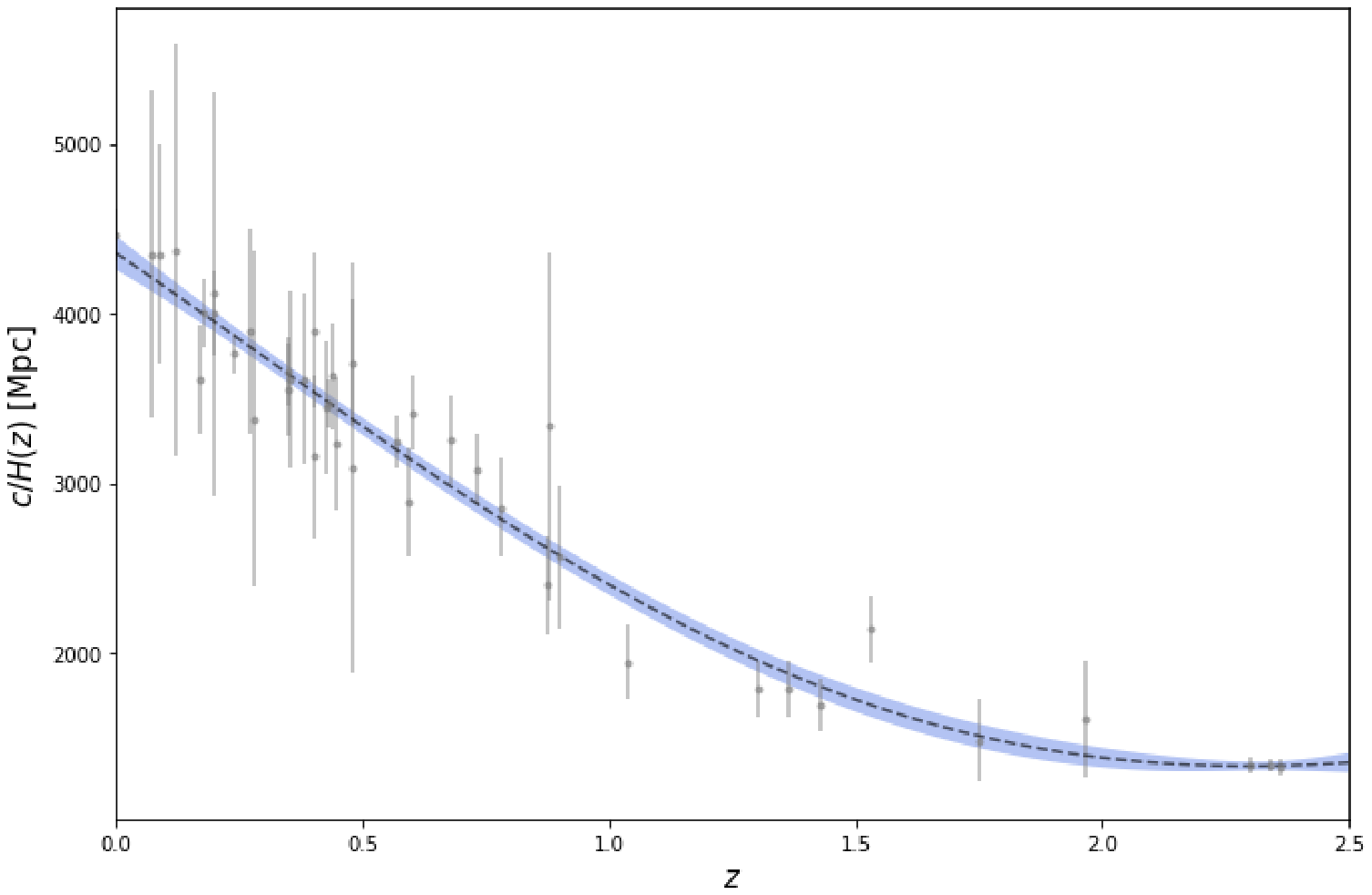}
\includegraphics[height=2.1in, width=3.1in]{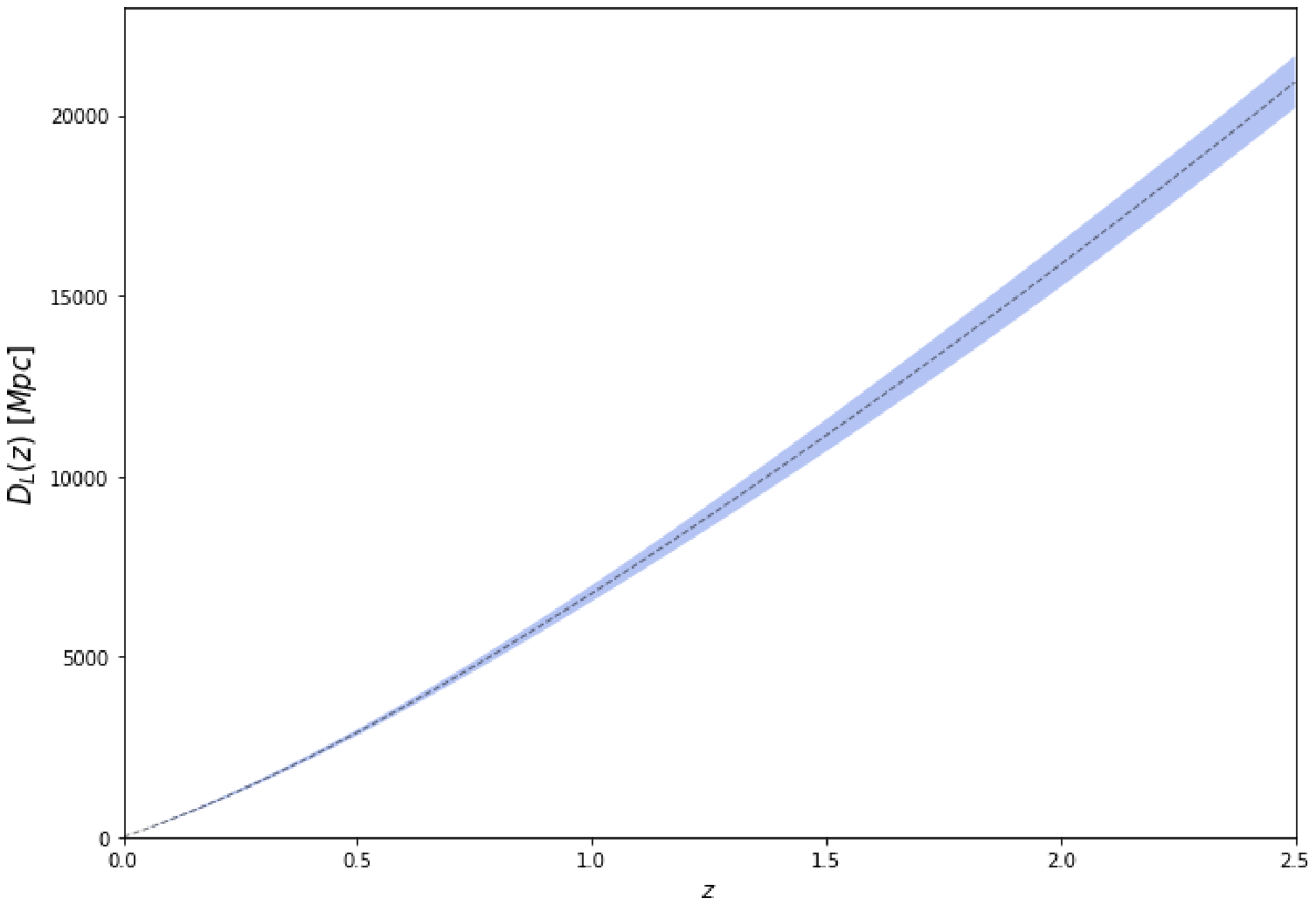}}
\caption{ Left panel: Recent measurements of Hubble parameter
measurements (black points) and the reconstruction of $c/H(z)$
function with the GP; Right panel: The corresponding reconstructed
luminosity distance $D_L(z)$ with the GP (given the covariance
matrix between the reconstructed $c/H(z)$ points). The blue region
represents the 1$\sigma$ confidence region.} \label{sb1}
\end{figure*}

\section{Observations}

It is only quite recently when reasonable catalogues of HII galaxies
and extragalactic HII regions containing more than 100 sources, with
spectroscopic as well as astrometric data are becoming available. In
this work, we have considered the current observations for 156 HII
objects compiled by \citet{Terlevich2015}. This data set was a
larger sample of sources than used by \citet{Siegel2005} or by
\citet{Plionis2011} and with more complete high-redshift data than
used by \citet{Melnick1988}.

On the one hand, the first complete sample from low-redshift
($0.01<z<0.2$) HII galaxies selected from the SDSS DR7 spectroscopic
catalogue is provided in \citep{Abazajian2009}. It consists of 128
local HII galaxies satisfying the following well-defined selection
criteria: 1) The lower limit of the equivalent widths of strongest
emission lines relative to the continuum is EW(H$\beta$)$>$50{\AA},
in order to guarantee the dominating contribution of a single young
starburst to the total luminosity, without the contamination of
underlying older population and older clusters
\citep{Dottori1981,DB1981,Melnick2000,Chavez2014,JunJie2016}); 2)
Extra objects with highly asymmetric emission lines should not be
included in the final sample \citep{Chavez2014}; 3) The upper limit
of the velocity dispersion is imposed as log$\sigma$(H$\beta$)$<$
1.8 km/s, in order to exclude rotationally supported systems and/or
objects with multiple young ionizing clusters contributing to the
total flux and affecting the line profiles \citep{Chavez2014}. When
applying the former two criteria to the full sample, 14 objects are
removed with the two cuts, while 7 more objects are excluded due to
their high velocity dispersion measurements. Therefore, we have the
``benchmark" catalog comprised of 107 local objects. On the other
hand, we also use a combined sample of 25 high-$z$ HII galaxies
covering the redshift range of $0.64\leq z\leq 2.33$ (taken from the
XShooter spectrograph at the Cassegrain focus of the ESO-VLT
\citep{Terlevich2015} and the literature
\citep{Erb2006a,Erb2006b,Masters2014}), as well as 24 giant HII
regions in nine nearby galaxies (taken from \citet{Chavez2012}),
which satisfy the well-defined observational selection criteria
listed above. See \citet{Melnick1987} for the measured velocity
dispersions and global integrated H$\beta$ fluxes with corresponding
extinction.

Full information about all 156 sources that remain after the
aforementioned selection can be found in Table 1 of
\citet{JunJie2016}, including source names, redshifts, categories,
integrated H$\beta$ flux, and corrected velocity dispersion. We
remark here that, the final sample covers the redshift range
$0<z<2.33$, which indicates its potential usefulness in cosmology at
high redshifts. From observational perspective, the reddening
corrected H$\beta$ flux is measured by fitting a single Gaussian to
the long-slit spectra \citep{Terlevich2015}, with the reddening
corrections derived from the published $E(B-V)$ using a standard
reddening curve \citep{Calzetti00}. The velocity dispersion inside
the aperture can also be derived from the spectroscopic data. More
specifically, one could obtain the velocity dispersion ($\sigma_0$)
and the corresponding 1$\sigma$ uncertainty from the full width at
half-maximum (FWHM) measurement of the H$\beta$ and
[OIII]$\lambda$5007 line, i.e.,
\begin{equation}
\sigma_0 \equiv \frac{FWHM}{2\sqrt{2\ln(2)}}.
\end{equation}
Following the strategy of \citet{JunJie2016}, the final corrected
velocity dispersion is defined as
\begin{equation}
\sigma = \sqrt{\sigma_0^{2} - \sigma_{th}^{2} - \sigma_i^{2} -
\sigma_{fs}^{2}},
\end{equation}
with the thermal broadening ($\sigma_{th}$), instrumental broadening
($\sigma_i$), and fine-structure broadening ($\sigma_{fs}$). See
\citet{Chavez2014} for more detailed discussion of the thermal and
instrumental broadening, while the fine-structure broadening is
taken as $\sigma_{fs}=$2.4 km/s, following the suggestion provided
in \citet{Gar2008}.

The test of the ``$L$--$\sigma$" relation of the standard candles
requires a statistically complete and well-characterized
(homogeneous) sample. Following the previous procedure applied to
compact radio sources \citep{Cao15c,Cao17a,Cao17b} and
galactic-scale strong lensing systems \citep{Cao16}, because our
list includes a wide class of HII objects at different redshift,
besides the full combined sample we will also consider separately
three sub-samples: high-$z$ HIIGx, local HIIGx, and GEHR.

\section{Methodology}

Following the phenomenological model first proposed in
\citet{Chavez2012} and later discussed in
\citet{Chavez2014,Terlevich2015}, the emission-line luminosity of a
source is related to its ionized gas velocity dispersion as
\begin{equation}
\log L ( \textrm{H}\beta) = \alpha \log \sigma(\textrm{H}\beta)+\kappa,
\end{equation}
where $\alpha$ is the constant slope parameter and $\kappa$
represents the logarithmic luminosity at $\log \sigma($H$\beta)=0$.
This is an empirical formula, whose scatter has been proved to be
very small that it can be effectively used as a luminosity indicator
in cosmology \citep{Chavez2012,Terlevich2015}. Meanwhile, The
H$\beta$ luminosity of the sources is estimated from their reddening
corrected flux density, which, assuming isotropic emission, reads
\begin{equation}
L(\textrm{H}\beta) = 4\pi D^{2}_L(z) F(\textrm{H}\beta),
\end{equation}
where $F(\textrm{H}\beta)$ is the reddening corrected H$\beta$ flux
and $D_L(z)$ is the luminosity distance at redshift $z$.

The combination of Eq.~(3)-(4) imply that if we could have a
reliable knowledge of cosmological distances at different redshifts,
then we would get stringent constraints on the range of parameters
$\alpha$ and $\kappa$ describing HII sources. Compared with the
previous procedure of simultaneously restricting ($\alpha,\kappa$)
with the cosmological parameter $\Omega_m$ (in the framework of
$\Lambda$CDM, XCDM and $Rh=ct$ cosmology) \citep{JunJie2016}, in
this work we try to place stringent constraints on the
``$L$--$\sigma$" relation in both the electromagnetic (EM) and
gravitational wave (GW) window, using luminosity distances covering
the HII redshift range derived in two cosmological model -
independent methods. Note that the strong degeneracies between
$\Omega_m$ and the two parameters characterizing the
``$L$--$\sigma$" relation, not only confirm that the cosmological
parameters are not independent of the nuisance parameters, but also
attest to the motivation of our calculation \citep{JunJie2016}.

In order to set limits on $\alpha$ and $\kappa$, we turn to two
catalogues of $D_L(z)$ separately by two different methods. In the
EM window, we will use luminosity distances derived in a
cosmological model-independent way from $H(z)$ measurements using
Gaussian processes (GP) \citep{Seikel12a}. As is well known,
assuming the FLRW metric of a flat universe, the angular diameter
distance can be written as
\begin{equation}
D_L(z)=(1+z)\int_0^z\frac{c\ dz}{H(z)},
\end{equation}
where $c$ is the speed of light and $H(z)$ is the Hubble parameter
at redshift $z$. The idea of cosmological application of GP
technique in general and with respect to $H(z)$ data can be traced
back to the paper of \citet{Holsclaw10} and then extensively applied
in more recent papers to test the cosmological parameters
\citep{Cao17a,Cao18}, spatial curvature of the Universe
\citep{Cao19,Qi19a}, and the speed of light at higher redshifts
\citep{Cao17b}. In this analysis, following the recent works of
\citet{Zheng19} inspired by Gaussian processes (GPs), we have
reconstructed the $c/H(z)$ function from the recent Hubble parameter
measurements including 41 data points from the galaxy differential
age method and 10 data points from the radial BAO size method, and
then derived $D_L(z)$ covering the redshift range of HII
observations \footnote{The Hubble constant $H_0=67.3$ km/s/Mpc from
the latest Planck CMB observations \citep{Planck18} is also taken
for distance reconstruction in our analysis.}. See
\citet{Qi18,Zheng19} for details and reference to the source papers.
The advantage of the Gaussian processes is, we do not need to assume
any parametrized model for $H(z)$ while reconstructing this function
from the data, which may provide more precise measurements of
angular diameter distances at a certain redshift. We use the
publicly available code called the GaPP (Gaussian Processes in
Python) \footnote{http://www.acgc.uct.ac.za/~seikel/GAPP/index.html}
reconstruct the profile of $H(z)$ function up to the redshifts
$z=2.5$, which can subsequently be used to reconstruct the
luminosity distance.

The GP method uses some attributes of a Gaussian distribution, i.e.,
the reconstructed function $f(z)$ follows a Gaussian distribution
with a mean value $\mu(z)$ and Gaussian error $\sigma(z)$ at each
point $z$. In this process, the values of the reconstructed function
evaluated at any two different points ($z$ and $\tilde{z}$) are
connected by a covariance function $k(z,\tilde{z})$, which depends
only on a set of hyperparameters ($\ell$ and $\sigma_f$). Compared
with the squared exponential covariance function widely used in the
previous studies \citep{Seikel12a,Seikel12b,Cai,Yang15,Qi19a}, we
take the Mat\'{e}rn ($\nu = 9/2$) form for the co-variance function
\begin{align}
k(z,\tilde z) = &~{\sigma_f}^2\exp\left( - \frac{{3\left| {z - \tilde z} \right|}}{\ell }\right) \nonumber \\
      &~~\times\Big[1 + \frac{{3\left| {z - \tilde z} \right|}}{\ell } + \frac{{27{{(z - \tilde z)}^2}}}{{7{\ell ^2}}} \nonumber \\
     &~~ + \frac{{18{{\left| {z - \tilde z} \right|}^3}}}{{7{\ell ^3}}} + \frac{{27{{(z - \tilde z)}^4}}}{{35{\ell
     ^4}}}\Big].
\end{align}
where $\sigma_f$ defines the overall amplitude of the correlation,
and $\ell$ gives a measure of the coherence length of the
correlation. The reliability of the reconstructed function can be
guaranteed by the fact that the hyper-parameters will be optimized
by the GP with the observational data sets, which furthermore
indicates that the reconstructed function is independent of the
initial hyper-parameter settings. In this analysis, an issue which
needs clarification is the achievable estimation of the $1\sigma$
confidence region for the reconstructed function $c/H(z)$. Note that
the $1\sigma$ confidence region depends on both the actual errors of
individual data points ($\sigma_{\frac{c}{H(z)}}$) and the product
$K_*K^{-1}K_*^T$. Here $K_*$ is the covariance matrix at redshift
$z_*$, which is calculated from the original $c/H(z)$ data at $z_i$
and the covariance matrix $k$:
\begin{equation}
K_*=[k(z_1,z_*),k(z_2,z_*),...,k(z_i,z_*)]\;.
\end{equation}
It should be pointed out that, when there is a large correlation
between the data ($K_*K^{-1}K_*^T>\sigma_f$), the dispersion at
point $z_i$ will be less than $\sigma_{\frac{c}{H(z)}}$ and the
reconstructed 1$\sigma$ regions will correspondingly become smaller.
More specifically, it was shown in the previous analysis
\citep{Seikel12a,Seikel12b} that the correlation between any two
points $z$ and $\tilde{z}$ will be large only when
$z-\tilde{z}<\sqrt{2}\ell$, which is clearly satisfied by the
current $H(z)$ data used in our analysis. Therefore, as can be seen
from the reconstructed results shown in Fig.~1, the GP estimated
$1\sigma$ confidence region is much smaller than the uncertainties
in the original $c/H(z)$ data. Such issue has been extensively
discussed in \citet{Seikel12a}. Using the reconstructed profile of
$c/H(z)$ function up to the redshifts $z\sim 2.5$, we are able to
reconstruct the luminosity distance $D_L(z)$ with the aforementioned
Gaussian processes. One should note that the error band should be
interpreted in a redshift by redshift sense and the covariances are
not visible in such a plot \citep{Seikel12a}. Following the
commonly-used procedure transforming $c/H(z)$ data into luminosity
distance \citep{Holanda12}, the $D_L(z)$ function can be calculated
by a usual simple trapezoidal rule (through Eq.~(5)). With the
standard error propagation formula, the error associated to the
$i_{th}$ redshift bin is given by $s_i=
\frac{1}{2}\left(\sigma_{\frac{c}{H(z_i)}}^2+\sigma_{\frac{c}{H(z_{i+1})}}^2\right)$,
where $\sigma_{\frac{c}{H(z)}}$ is the error of the $c/H(z)$ data
reconstructed from Gaussian processes. However, it should be noted
that the constructed luminosity distances are correlated, since all
of the derived $s_i$ are statistically dependent on each other
\citep{Liao15}. More specifically, the $c/H(z)$ data are Gaussian
Process reconstructed, following a multidimensional Gaussian
distribution with the covariance matrix (through Eq.~(7)). Hence,
the uncertainty of the luminosity distance corresponding to certain
redshift $z$ should include statistical uncertainties and the
covariances between every $c/H(z)$ pair among the total data. That
is,
\begin{equation}
\sigma^2_{D_{L,H(z)}}=\frac{(\Delta z)^2}{2}\left[\sum_{i=1}^{n}s_i+
\sum_{i=2}^{n}\sum_{j=1}^{i-1} {\rm Cov} \left(\frac{c}{H(z_i)},
\frac{c}{H(z_j)}\right)\right],
\end{equation}
where $\Delta z$ is the length of the redshift bin, while $\rm{Cov}$
denotes the covariance matrix for a set of reconstructed $c/H(z)$
points given by Eq.~(7). The results are also shown in Fig.~1, where
the reconstructed $D_{L,H(z)}$ function with corresponding 1$\sigma$
uncertainty strip are displayed. Distance reconstruction with the
Hubble parameter measurements is denoted as ``Cosmology-independent
method I".

\begin{table*}
\caption{\label{tab:result} Summary of the constraints on the
``$L$--$\sigma$" relation parameters obtained with the full sample
and three sub-samples (see text for definitions).}
\begin{center}
\begin{tabular}{l|l|llllllll}\hline\hline
Sample (Calibration method+Cosmology)        & $\alpha$ & $\kappa$  \\
\hline Full sample (Cosmology-independent method I)  & $\alpha=5.10\pm 0.10$ & $\kappa=33.12\pm 0.15$   \\
Full sample (Cosmology-independent method II+Planck) & $\alpha=5.13\pm 0.08$ & $\kappa=33.06\pm 0.13$   \\
Full sample (Cosmology-independent method II+WAMP9) & $\alpha=5.17\pm 0.09$ & $\kappa=32.86\pm 0.12$   \\
\hline

High-$z$ HIIGx (Cosmology-independent method I) & $\alpha=5.18\pm 0.65$ & $\kappa=33.00\pm 1.13$   \\
Local HIIGx (Cosmology-independent method I)  & $\alpha= 4.88\pm 0.15$ & $\kappa=33.48\pm 0.22$   \\
GEHR (Cosmology-independent method I) & $\alpha= 5.77\pm 0.52$ & $\kappa=32.25\pm 0.62$   \\
\hline

High-$z$ HIIGx (Cosmology-independent method II) & $\alpha=5.33\pm 0.65$ & $\kappa=32.70\pm 1.13$   \\
Local HIIGx (Cosmology-independent method II)  & $\alpha= 4.93\pm 0.14$ & $\kappa=33.39\pm 0.22$   \\
GEHR (Cosmology-independent method II) & $\alpha= 5.81\pm 0.50$ & $\kappa=32.19\pm 0.61$   \\

\hline\hline
\end{tabular}
\end{center}
\end{table*}

\begin{figure}
\begin{center}
\includegraphics[width=1.0\linewidth]{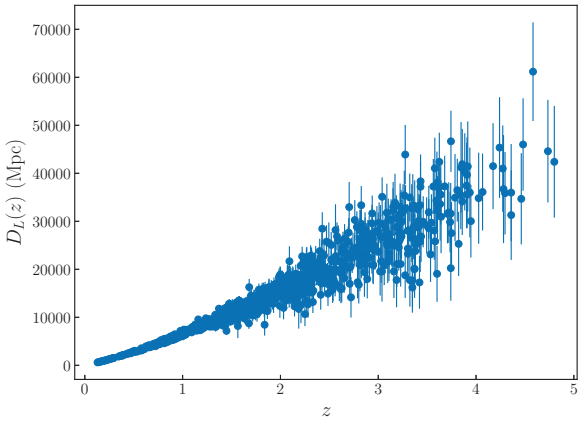}
\end{center}
\caption{The luminosity distance measurements from 1000 GW events
detected by ET.}
\end{figure}

In the GW window, we turn to the simulated data of gravitational
waves from the third-generation gravitational wave detector, which
can be considered as standard siren to provide the information of
luminosity distance. Gravitational waves provide us with a
completely new means of observation and are also a promising probe
for cosmology. It is well known that the detection of gravitational
waves (GW) from the merger of double compact object (DCO)
\citep{Abbott16,Abbott17} has opened the new era of GW astronomy.
The original idea of using the waveform signal to directly measure
the luminosity distance $D_L$ to the GW sources can be traced back
to the paper of \citet{Schutz86}, which indicates the inspiraling
and merging compact binaries consisting of neutron stars (NSs) and
black holes (BHs), can be used to constrain the Hubble constant by
combining the redshift information of source. Therefore,
gravitational wave signals from the merger of DCOs are put forward
as distance indicators and are called standard sirens
\citep{Dalal2006,Taylor12,Zhao11,Cai15,Cai17}. If we can locate the
host galaxy by means of EM counterparts, redshift information of the
GW source can be easily obtained. In this paper we simulate GW
events based on the Einstein Telescope, the third generation of the
ground-based GW detector \citep{ET}. Although only a few GW events
have been detected by the current advanced ground-based detectors
(i.e., the advanced LIGO and Virgo detectors), ET will expand the
detection space by three orders of magnitude, and thus can detect
much more GW events \citep{Cai15,Cai17}. In this paper, we carry out
a Monte Carlo simulation of the GW signals of NS-NS and NS-BH
systems with high signal to noise ratio (SNR), based on future
observations from the third generation technology (the ``xylophone"
configuration) \citep{Cai15}. The specific steps to simulate the
mock data is similar with that used in \citet{Qi19b,Qi19c}.
Concerning the error strategy, the combined signal-to-noise ratio
(SNR) for the network not only helps us confirm the detection of GW
with $\rho_{net}>8$, the SNR threshold currently used by LIGO/Virgo
network, but also contribute to the error on the luminosity distance
as $\sigma^{inst}_{D_{L,GW}}\simeq\frac{2D_{L,GW}}{\rho}$
\citep{Zhao11}. Meanwhile,the lensing uncertainty caused by the weak
lensing is also taken into consideration, which is modeled as
$\sigma^{lens}_{D_{L,GW}}/D_{L,GW}=0.05z$
\citep{Sathyaprakash2010,Li2015}. Therefore, the distance precision
per GW is taken as
\begin{eqnarray}
\sigma_{D_{L,GW}}&=&\sqrt{(\sigma_{D_{L,GW}}^{\rm inst})^2+(\sigma_{D_{L,GW}}^{\rm lens})^2} \nonumber\\
            &=&\sqrt{\left(\frac{2D_{L,GW}}{\rho}\right)^2+(0.05z D_{L,GW})^2}.
\label{sigmadl}
\end{eqnarray}
In this paper, we take the flat $\Lambda$CDM universe as our
fiducial model in the simulation. The matter density parameter
$\Omega_m=0.315$ and the Hubble constant $H_0=67.3$ km/s/Mpc from
the latest Planck CMB observations \citep{Planck18} is taken for
Monte Carlo simulations in our analysis. Following the redshift
distribution of GW sources taken as \citep{Sathyaprakash2010} and
assuming the luminosity distance measurements obey the Gaussian
distribution, the simulated 1000 GW events used for statistical
analysis in the next section are shown in Fig.~2. In our analysis,
in order to get $D_L$ at the redshift of HII galaxy, we have
employed the Gaussian Processes (GPs) to reconstruct the function
$D_{L,GW}(z)$ and its corresponding $1\sigma$ uncertainty
$\sigma_{D_{L,GW}}$. Distance reconstruction with the simulated GW
sample is denoted as ``Cosmology-independent method II".

Now, from the observational point of view, in the framework of
``$L$--$\sigma$" relation, the observed distance modulus of an HII
object is
\begin{equation}
\mu_{obs} = 2.5[\kappa +\alpha \log \sigma(\textrm{H}\beta)-\log
F(\textrm{H}\beta)]-100.2,
\end{equation}
with the corresponding error $\sigma_{\mu_{obs}}$ expressed as
$\sigma_{\mu_{obs}}= \sqrt{(2.5\alpha \sigma_{\log
\sigma})^{2}+(2.5\sigma_{\log F})^{2}}$. Here $\sigma_{\log \sigma}
$ and $\sigma_{\log F}$ represent the standard errors of the
reddening corrected H$\beta$ flux ($\log \sigma($H$\beta)$) and the
corrected velocity dispersion ($\log F($H$\beta)$). For each HII
galaxy, the reconstructed distance modulus $\mu_{th}$ can be
calculated from the measured redshift $z$ by the definition
\begin{equation}
\mu_{th} \equiv 5\log\left[\frac{D_L(z)}{Mpc}\right]+25,
\end{equation}
where $D_L(z)$ is the cosmology-dependent luminosity distance
obtained through ``Cosmology-independent method I" and
``Cosmology-independent method II". The propagated uncertainty of
$\mu_{th}$ is given by $\sigma_{\mu_{th}}=\frac{5\sigma_{D_L}}{D_L
\ln10}$. We determine the parameters ($\alpha$ and $\kappa$)
characterizing HII objects by minimizing the $\chi^2$ objective
function
\begin{equation}
\chi^{2}(\alpha, \kappa) = \sum_i \frac{(\mu_{obs}(z_i;
\alpha,\kappa )-\mu_{th}(z_i))^{2}}{\sigma_{\mu,i}^{2}}
\end{equation}
and the corresponding statistical error is given by
\begin{equation}
\sigma^{2}= (2.5\alpha \sigma_{\log \sigma})^{2}+(2.5\sigma_{\log
F})^{2}+(\frac{5\sigma_{D_L}}{D_L \ln10})^{2}.
\end{equation}
Note that the observational statistical uncertainty for the ith data
point and the uncertainty for the reconstructed distance modulus are
both included. Then using the Markov Chain Monte Carlo technique
available within CosmoMC package \citep{Lewis02}, we preform Monte
Carlo simulations of the posterior likelihood ${\cal L} \sim \exp{(-
\chi^2 / 2)}$ and apply a public python package ``triangle.py" from
\citet{Foreman13} to plot our constraint contours.

\begin{figure}
\begin{center}
\includegraphics[width=1.0\linewidth]{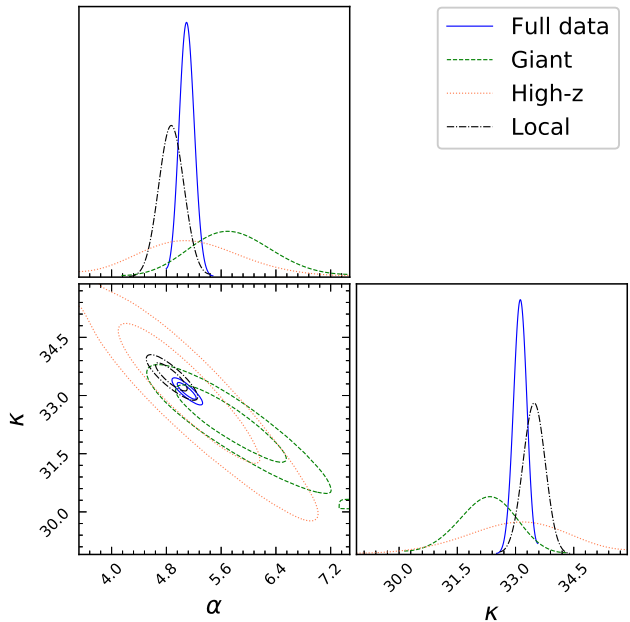}
\end{center}
\caption{ Constraints on HII parameters obtained from the full
sample and three sub-samples (high-$z$ HIIGx, local HIIGx, and
GEHR), based on the $D_L(z)$ function reconstructed from current
$H(z)$ data (``Cosmology-independent method I"). }
\end{figure}

\begin{figure}
\begin{center}
\includegraphics[width=1.0\linewidth]{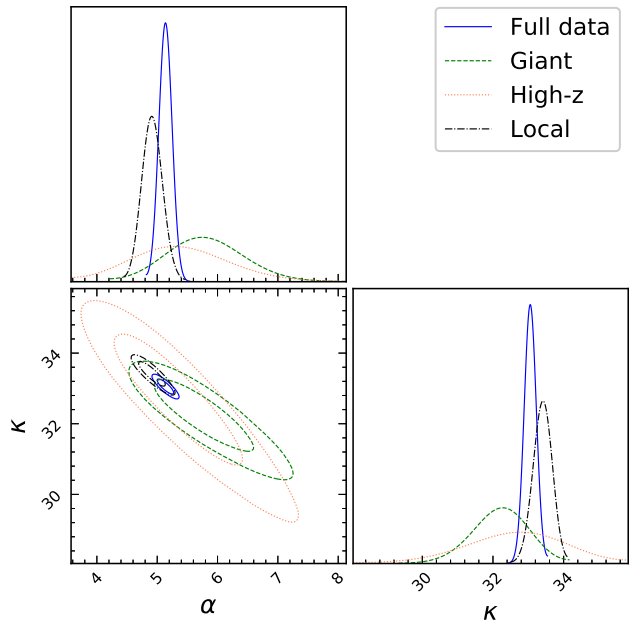}
\end{center}
\caption{Constraints on HII parameters obtained from the full sample
and three sub-samples (high-$z$ HIIGx, local HIIGx, and GEHR), based
on the $D_L(z)$ measurements from future simulated GW data
(``Cosmology-independent method II").}
\end{figure}

\begin{figure*}
\begin{center}
\includegraphics[width=0.7\linewidth]{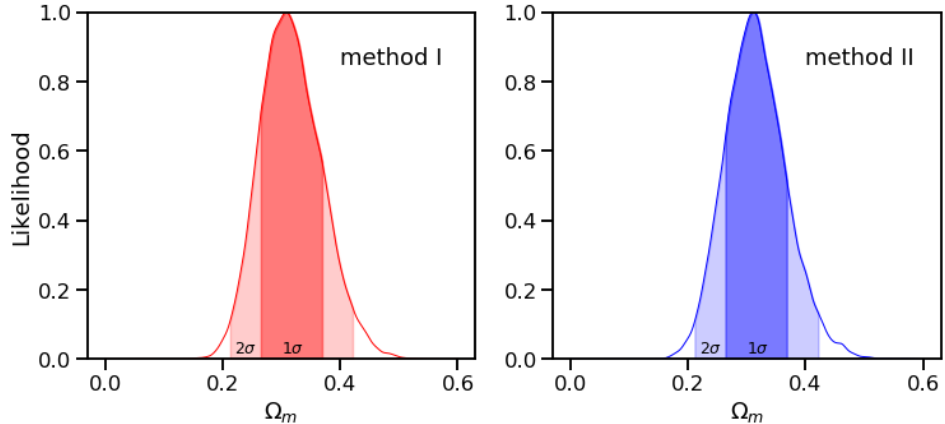}
\end{center}
\caption{Cosmological fits on the flat $\Lambda$CDM model obtained
from the full sample, based on the corrected ``$L$--$\sigma$"
relation with the current $H(z)$ data (left panel) and future
simulated GW data (right panel).}
\end{figure*}

\section{Results and discussions}

In this section, we focus our attention on the constraints on the
parameters ($\alpha$ and $\kappa$) obtained from different samples,
i.e., the full $N=156$ sample, as well as three sub-samples
determined from high-$z$ HIIGx, local HIIGx, and GEHR. The results
are summarized in Table I. The graphic representations of the
probability distribution of $\alpha$ and $\kappa$ are presented in
Figs. 3-4, in which one can see the 1-D distributions for each
parameter and 1$\sigma$, 2$\sigma$ contours for the joint
distribution.

To start with, by applying the above mentioned $\chi^2$-minimization
procedure to the distance reconstruction with the Hubble parameter
measurements (``Cosmology-independent method I"), we obtain the
results shown in Fig.~3. Performing fits on the full data comprising
156 objects, we obtain the following best-fit values and
corresponding 1$\sigma$ uncertainties (68.3\% confidence level):
\begin{eqnarray}
&& \alpha=5.10\pm 0.10, \nonumber\\
&& \kappa=33.12\pm 0.15. \nonumber
\end{eqnarray}
Marginalized 1$\sigma$ and 2$\sigma$ contours of each parameter
obtained are shown in Fig.~3. It is obvious that, the full sample
analysis has also yielded improved constraints on the meaningful
physical parameters: $\alpha$ and $\kappa$. More importantly, we
find that our constraints on the two parameters with
``Cosmology-independent method I" are very different from those
obtained in the framework of different cosmologies. For instance,
some researchers \citep{JunJie2016} have previously derived a fit to
the flat $\Lambda$CDM, XCDM and $Rh=ct$ cosmology, with the
optimized parameter values for the $\alpha$ parameter:
$\alpha=4.89\pm0.09$, $\alpha=4.87\pm0.10$ and $\alpha=4.86\pm0.08$,
which disagrees with our results at 68.3\% C. L. Therefore, the
values of the two best-fit parameters of the phenomenological
formula obtained in our analysis, if confirmed by future
investigation of HII observations, will offer additional constraints
for cosmological tests based on ``$L$--$\sigma$" relation of
extragalactic sources.

In Table 1 and Fig.~3, we show the results of fitting the two
parameters, $\alpha$ and $\kappa$, on three sub-samples described in
Section II. It is interesting to note that the ranges of $\alpha$
and $\kappa$ for local HII galaxies ($\alpha= 4.88\pm 0.15$,
$\kappa=33.48\pm 0.22$) are marginally close to estimates obtained
from High-$z$ HII galaxies ($\alpha=5.18\pm 0.65$, $\kappa=33.00\pm
1.13$). On the other hand, the constrained results for giant
extragalactic HII regions (GEHR), which constitute the most
important part of our full HII sample are particularly interesting.
Namely, one can clearly see that the best-fit values of the two
parameters for this population, $\alpha= 5.77\pm 0.52$ and
$\kappa=32.25\pm 0.62$, are significantly different from the
corresponding quantities of HII galaxies. Substantial distinction
between $\alpha$ and $\kappa$ parameters exists for the two
sub-populations (GEHR and HIIGx) is more clear when the 1$\sigma$
uncertainties are taken into consideration. Consequently, our
results indicate the different ``$L$--$\sigma$" relation of HII
regions acting as standard candles.

Then one issue which might be raised is the choice of the $D_A(z)$
function reconstructed from current $H(z)$ data in the course our
estimation of $\alpha$ and $\kappa$. Therefore, we have undertaken a
similar analysis with the second model-independent approach, the
simulated data of gravitational waves from the third-generation
gravitational wave detector \footnote{Note that in the second
approach with simulated GW data, we pay more attention to
demonstrating the improvements that future GW measurements could
provide, concerning the calibration of the ``$L$--$\sigma$"
relation.}. In this case, performing fits on the full data set, the
68.3\% confidence level uncertainties on the three model parameters
are:
\begin{eqnarray}
&& \alpha=5.13\pm 0.08, \nonumber\\
&& \kappa=33.06\pm 0.13. \nonumber
\end{eqnarray}
Fig.~4 shows these constraints in the parameter space of $\alpha$
and $\kappa$. Comparing constraints based on the two
model-independent methods, we see that confidence regions of
$\alpha$ and $\kappa$ are well overlapped with each other; hence our
results and discussions presented above are robust. This tendency
could also be found in fits performed on three sub-samples with
local HII galaxies, high-$z$ HII galaxies, and giant extragalactic
HII regions. From the results displayed in Fig.~4, one can find the
obtained value of $\alpha$ and $\kappa$ from our sub-sample with
giant extragalactic HII regions, whose confidence contours in the
($\alpha$, $\kappa$) parameter plane differ from the other two
remaining samples. More specifically, in the framework of the
``$L$--$\sigma$" relation for giant extragalactic HII regions, a
lower value of the slope parameter and a higher value of the
logarithmic luminosity at $\log \sigma($H$\beta)=0$ is revealed and
supported by our analysis. We must keep in mind that similarity or
difference in ($\alpha$, $\kappa$) parameters for HII observations
with different types of optical counterparts might reveal similar or
different physical processes governing the H$\beta$ emission in GEHR
and HIIGx. To some extent, our results imply the need of treating
these classes of HII observations separately in future cosmological
studies.

The second issue which needs clarification is the fiducial cosmology
used in our GW simulation, i.e., the consistency between the
luminosity distance coming from GP reconstructed $H(z)$ and the
simulated GW standard siren should be fully tested. In order to
explore the potential systematics caused by different priors of
cosmological parameters, besides assuming a flat $\Lambda$CDM model
with parameters coming from Planck 2018 observations, we also
consider the WMAP nine year results (WMAP9) for comparison, in which
the matter density parameter and the Hubble constant are
respectively taken as $\Omega_m=0.279$ and $H_0=70.0$ km/s/Mpc
\citep{WMAP13}. In this case, the full data set provides the best
fit on the ``$L$--$\sigma$" relation as
\begin{eqnarray}
&& \alpha=5.17\pm 0.09, \nonumber\\
&& \kappa=32.86\pm 0.12. \nonumber
\end{eqnarray}
Comparing constraints based on Planck and WMAP9 observations shown
in Table 1, one could see that confidence regions of $\alpha$ and
$\kappa$ are almost the same. We remark here that, considering that
the WMAP9 and Planck data are consistent with the accuracy
sufficient to the comparison with the ``$L$--$\sigma$" relation, it
is not surprising that the regression results of the
``$L$--$\sigma$" relation in combination with WMAP and Planck are
compatible in the framework of $\Lambda$CDM cosmology
\citep{Cao15c}.

Having performed cosmological-model-independent analysis, we can
also investigate cosmological implications of the distance modulus
of 156 HII measurements by taking the corrected ``$L$--$\sigma$"
relation into consideration. In this analysis we focus on the
$\Lambda$CDM model when spatial flatness of the FLRW metric is
assumed, which is strongly indicated by the location of the first
acoustic peak in the CMBR \citep{Planck18} and also independently
supported by the quasar data at $z\sim 3.0$ as demonstrated in
\citep{Cao19}. The Friedmann equation is
\begin{equation}
H^2 = H_0^2 \ [\Omega_m(1+z)^3+1-\Omega_m] \, ,
\end{equation}
where $\Omega_m$ parameterizes the density of matter (both baryonic
and non-baryonic components) in the Universe. For the flat
$\Lambda$CDM model, different from the methods used in
\citet{JunJie2016}, we examine the probability distributions of
$\Omega_m$ by considering the best-fitted $\alpha$ and $\beta$
parameters (with their 1$\sigma$ uncertainties) obtained from the
previous model-independent tests. Fitting the $\Lambda$CDM model to
the full sample with the corrected ``$L$--$\sigma$" relation, one is
able to get the observational constraint on the matter density
parameter as $\Omega_{m}=0.314\pm0.054$ (calibrated with standard
clocks in EM domain) and $\Omega_{m}=0.311\pm0.049$ (calibrated with
standard sirens in GW domain). The results are shown in Fig.~5. On
the one hand, one may observe that the results obtained from the
combined HII sample are well consistent with the fit based on the
full-mission Planck observations of temperature and polarization
anisotropies of the CMB radiation \citep{Planck18}, as well as a
newly compiled data set of mas compact radio sources representing
intermediate-luminosity quasars covering the redshift range
$0.5<z<2.8$ \citep{Cao17a,Cao17b,Li17,Xu18}. On the other hand, our
results strongly suggest that the dynamical properties of HII
galaxies may significantly impact the likelihood distributions of
$\Omega_m$ and thus constraints on the properties of dark energy.
This conclusion is strengthened by the comparison of our
cosmological fits from the recalibrated ``$L$--$\sigma$" relation
through our cosmological-model-independent tests and those based on
a specific cosmological scenario \citep{JunJie2016}. Therefore,
although the constraints resulting from this analysis are marginally
consistent with the previous works, our results based on a
cosmological-model-independent check (especially
``Cosmology-independent method I") could be useful as hints for
priors on $\alpha$ and $\kappa$ parameters in future cosmological
studies using HII observations.

\section{Conclusion}

In this paper, we explored the properties of a sample of 156 HII
galaxies (HIIGx) and giant extragalactic HII regions (GEHR) with
measured flux density and turbulent gas velocity. The
``$L$--$\sigma$" relation of these standard candles is usually
parameterized as $\log L ($H$\beta) = \alpha \log
\sigma($H$\beta)+\kappa$. Using the cosmological distances
reconstructed through two new cosmology-independent methods, we
investigate the correlation between the emission-line luminosity $L$
and ionized-gas velocity dispersion $\sigma$. The method is based on
non-parametric reconstruction using the measurements of Hubble
parameters from cosmic clocks, as well as the the simulated data of
gravitational waves from the third-generation gravitational wave
detector (the Einstein Telescope, ET), which can be considered as
standard sirens. Moreover, we have also investigate cosmological
implications of the distance modulus of 156 HII measurements by
taking the corrected ``$L$--$\sigma$" relation into consideration,
which encourages us to probe cosmological parameters beyond the
current reach of Type Ia supernovae. Here we summarize our main
conclusions in more detail:

\begin{itemize}

\item In the full sample, we find that measurements of HIIGx and
GEHR provide tighter estimates of the ``$L$--$\sigma$" relation
parameters. Performing fits on the full data comprising 156 objects,
we obtain the following best-fit values and corresponding 1$\sigma$
uncertainties (68.3\% confidence level): $\alpha=5.10\pm 0.10,
\kappa=33.12\pm 0.15$ (calibrated with standard clocks in EM domain)
and $\alpha=5.13\pm 0.08, \kappa=33.06\pm 0.13$ (calibrated with
standard sirens in GW domain). We have also explored the potential
systematics caused by different priors of cosmological parameters in
GW simulation. In the framework of a flat $\Lambda$CDM model with
parameters coming from WMAP9, the full data set provides the best
fit on the ``$L$--$\sigma$" relation: $\alpha=5.17\pm 0.09$ and
$\kappa=32.86\pm 0.12$ (calibrated with standard sirens in GW
domain). More importantly, our constraints on the two parameters
with two new cosmology-independent methods are very different from
those obtained in the framework of different cosmologies.

\item Furthermore, we divide the full sample into three different
sub-samples according to their optical counterparts. It turns out
that the ranges of $\alpha$ and $\kappa$ for local HII galaxies are
marginally close to estimates obtained from High-$z$ HII galaxies.
The best-fit values for giant extragalactic HII regions (GEHR) are
significantly different from the corresponding quantities of HII
galaxies. Substantial distinction between $\alpha$ and $\kappa$
parameters exists for the two sub-populations (GEHR and HIIGx) is
more clear when the 1$\sigma$ uncertainties are taken into
consideration. Consequently, closeness or difference of parameter
values for different types of counterparts indicate the similar or
different ``$L$--$\sigma$" relation of HII regions acting as
standard candles, as well as the existence of possible similar or
different physical processes governing the H$\beta$ emission in GEHR
and HIIGx.

\item Fitting the $\Lambda$CDM model to the full sample with the corrected
``$L$--$\sigma$" relation, one is able to get the observational
constraint on the matter density parameter as
$\Omega_{m}=0.314\pm0.054$ and $\Omega_{m}=0.311\pm0.049$, which is
inconsistent with the previous results obtained on the same sample
but agrees very well with other recent astrophysical measurements
including Planck observations. Therefore, it is strongly suggested
that reliable knowledge of the dynamical properties of HII galaxies
may significantly impact the constraints on relevant cosmological
parameters. The values of the two best-fit parameters of the
``$L$--$\sigma$" relation obtained in our analysis, if confirmed by
future investigation of HII observations, will offer additional
constraints for cosmological tests based on extragalactic sources.

\item As a final remark, we point out that the sample discussed in
this paper is based on HII objects discovered in different surveys.
Our analysis potentially may suffer from systematics stemming from
this inhomogeneity. Therefore, we may expect more vigorous and
convincing constraints on the dynamical properties of HII galaxies
within the coming years with more precise data, especially a larger
sample of high-z HIIGx observed by the current facilities such as
the K-band Multi Object Spectrograph at the Very Large Telescope
\citep{Terlevich2015}.

\end{itemize}

\vspace{0.5cm}

We are grateful to Jingzhao Qi for helpful discussions. This work
was supported by National Key R\&D Program of China No.
2017YFA0402600, the National Natural Science Foundation of China
under Grants Nos. 11503001, 11690023, 11373014, and 11633001, the
Strategic Priority Research Program of the Chinese Academy of
Sciences, Grant No. XDB23000000, the Interdiscipline Research Funds
of Beijing Normal University, and the Opening Project of Key
Laboratory of Computational Astrophysics, National Astronomical
Observatories, Chinese Academy of Sciences.

\end{document}